# A Note on Half-Quantized Conductance Plateau of Chiral Majorana in Quantum Anomalous Hall Insulator and Superconductor Structures

Peng Zhang[1], Lei Pan[1], Gen Yin[1], Qing Lin He[2], and Kang L. Wang[1]*

[1]Department of Electrical and Computer Engineering and Department of Physics and Astronomy, University of California, Los Angeles, CA 90095, USA.

[2]International Center for Quantum Materials, School of Physics, Peking University, Beijing 100871, China

*Correspond to K. L. Wang: wang@ee.ucla.edu; klwang@ucla.edu

Recently, there was a letter [1] posted on arxiv.org/abs/1904.06463 by M. Kayyalha *et al.* reporting a transport study on a quantum anomalous Hall insulator (QAHI) – superconductor (SC) structure. In their arXiv letter, the authors used the device structure similar to that used in our previous report [2], with different substrate and details. Yet in their devices, the two-terminal conductance ($\sigma_{12}$) was found to always near but below ~0.5$e^2$/$h$ (half-plateaus), and could not go above this value at the quantum anomalous Hall temperature over a wide range of external magnetic field, *i.e.* from 0 to ~ 3 T as shown in Fig. 3 (a) and 4 of [1]. The authors attributed their observation to the shunting effect of two quantum anomalous Hall edges by the SC layer. Here, we would like to point out that the half-plateaus observed in their arXiv letter are, in fact, different from those observed from our previous results [2], in which the half-plateaus occur near the magnetization reversal regions (Fig. 1 of this note, reproduced from Ref. [2]).

In clarifying and addressing this issue, we present additional magneto-transport data for our similar QAHI-SC samples fabricated after Ref. [2] was published. The results taken from these new samples are shown in Fig. 2 of this note. It can be clearly seen that the half-plateaus are well developed before the magnetization reversal regions, again validating our previous results. Furthermore, by varying the conditions of samples as well as the materials properties, the widths of the half-plateaus can be adjusted. Our studies and our results on the new structures not only demonstrate that the appearance of the half-plateaus is reproducible over different devices, but also further point out that the QAHI-SC coupling, the details of processing, and material properties play important roles in obtaining the half-plateaus. More investigative experiments continue.

Acknowledgment

We thank Professors Kai Liu, Allan MacDonald, Xiaogang Wen, Jay Sau, Liang Fu, Wei-Li Lee, and Biao Lian for helpful discussions; Professor Ni Ni of UCLA and Professor Jing Shi of UC-Riverside for the use of their dilution refrigerators.

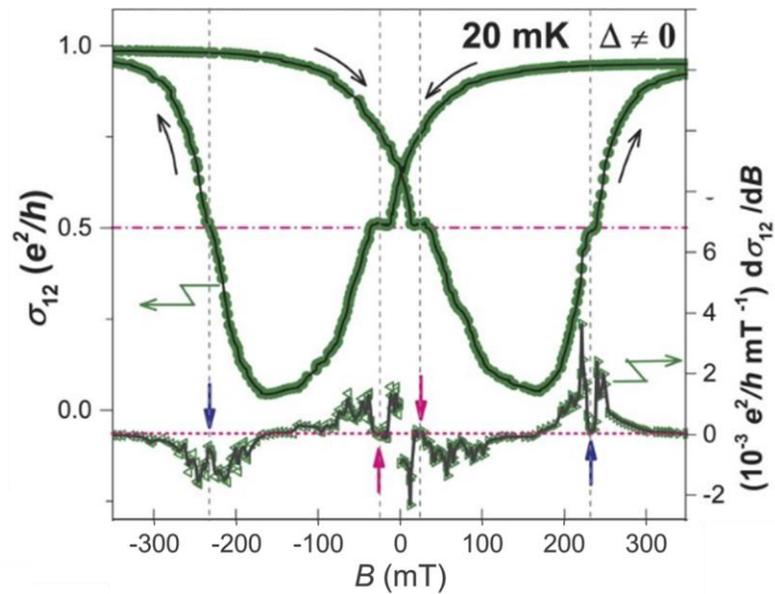

Fig. 1. $\sigma_{12}$ as a function of the magnetic field. Adopted from Fig. 2C in Ref. [2].

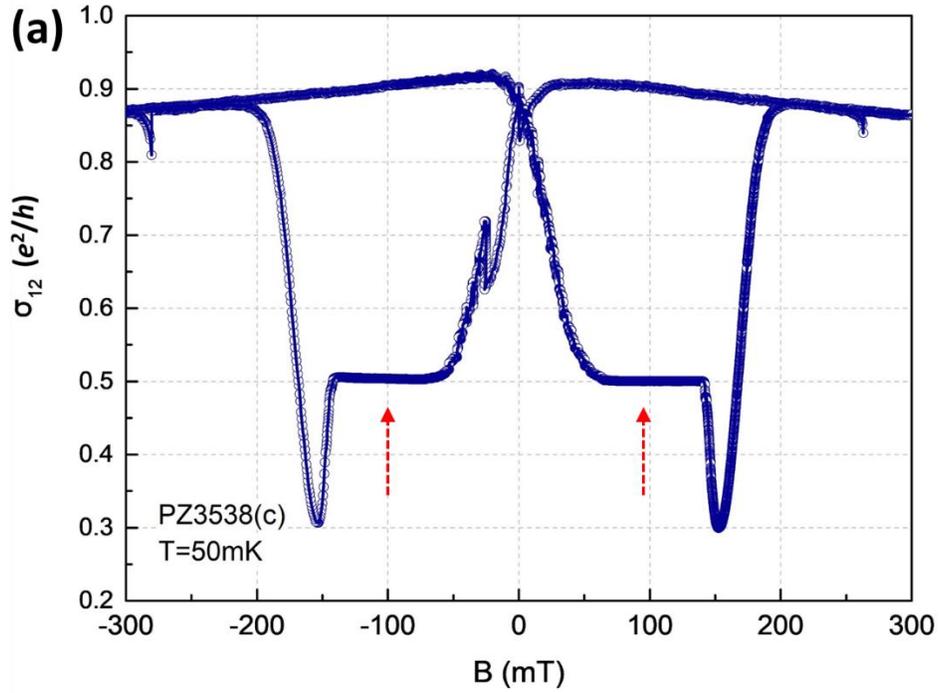

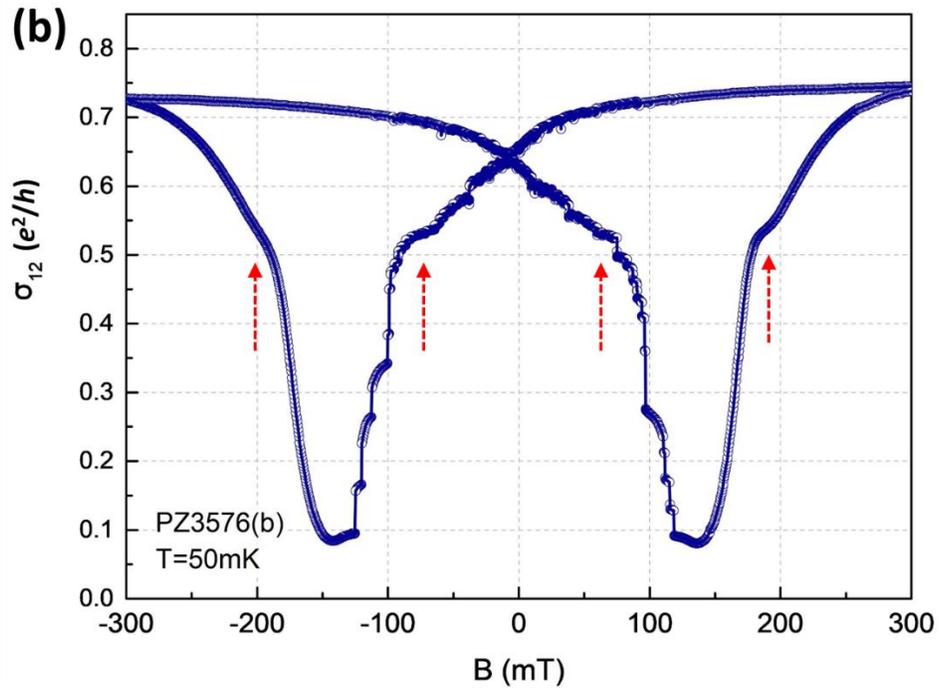

Fig. 2. $\sigma_{12}$ as a function of the magnetic field in two new devices directly extracted from the raw data. By controlling the fabrication process and the material properties, the widths of the half-plateaus (red arrows) can be adjusted as shown in (a) and (b). Data taken at 50 mK.